\documentclass[singlecolumn,showpacs,preprintnumbers,amsmath,amssymb,floatfix]{revtex4}
\usepackage{graphicx}% Include figure files
\usepackage{dcolumn}% Align table columns on decimal point
\usepackage{bm}% bold math

\begin{document}
%\preprint{APS/123-QED}

\title{Systematic study of the fusion barriers using different proximity-type potentials for $N = Z$ colliding nuclei: New extensions}
% Force line breaks with \\

\author{Ishwar Dutt }
%\altaffiliation [Also at ]{Physics Department, XYZ University.}%Lines break automatically or can be forced with \\
\author{Rajeev K. Puri}%
\email{rkpuri@pu.ac.in; drrkpuri@gmail.com}
\affiliation{Department of Physics, Panjab University, Chandigarh
160 014, India}

%\affiliation{

\date{\today}% It is always \today, today,

\begin{abstract}

By using 14 different versions and parametrizations of a proximity
potential and two new versions of the potential proposed in this
paper, we perform a comparative study of fusion barriers by
studying 26 symmetric reactions. The mass asymmetry $\eta_{A}$ =
$\left(\frac{A_{2}-A_{1}}{A_{2}+A_{1}}\right),$ however, is very
large. Our detailed investigation reveals that most of the
proximity potentials reproduce experimental data within $\pm$8\%
on the average. A comparison of fusion cross sections indicates
that Bass 80, AW 95, and Denisov DP potentials have a better edge
than other potentials. We also propose new versions of the
proximity potential as well as Denisov parametrized potential.
These new versions improve the agreement with the data.
\end{abstract}

\pacs{25.70.Jj,24.10.-i.}

% PACS, the Physics and Astronomy % Classification Scheme.
\maketitle

%break was forced \lowercase{via} \textbackslash\textbackslash}
\section{\label{intro}Introduction}
 The study of fusion barriers and fusion
cross sections has received renewed attention in recent
years~\cite{siwek04,gr09,rkp1,wang06,canto09}. This renewed
interest is caused by the efforts of low and intermediate energies
in order to understand the fusion mechanism and, subsequently, the
nucleus-nucleus interactions in nuclear physics. This is further
boosted by the availability of radioactive-ion beams involving
various reactions~\cite{canto09}.

 In recent years, large numbers of models depending on the vast variety
of assumptions have been
proposed~\cite{siwek04,rkp1,wang06,ms2000,blocki77,wr94,deni02,aw95}.
One set of such theoretical models is based on the microscopic
picture in which one starts from two- and three-body effective
interactions and calculates the ion-ion potential
~\cite{rkp1,wang06,deni02}. Another class of the models takes the
gross macroscopic picture into account~\cite{bass73,bass77}.
Because of the recent precise measurements of the fusion cross
sections, the job of theoretical models has become much more
complex. Among different theoretical models, the proximity
potential enjoys special status~\cite{ms2000,blocki77}. All
proximity potentials are based on the proximity force theorem,
according to which, the nuclear part of the interaction potential
can be written as the product of a factor depending on the mean
curvature of the interaction surface and a universal function
(depending on the separation distance) and  is independent of the
masses of colliding nuclei. This concept did introduce a great
amount of simplification in nuclear potential
studies~\cite{wang06,deni02}. Several refinements and
modifications have been proposed in the recent past concerning
over the original proximity potential to remove the gray part of
the potential~\cite{ms2000}. This observation demands a careful
and systematic study of the heavy-ion fusion process using all
such potentials that are remodeled and parametrized within the
proximity concept.
\par
Here, we will concentrate on the symmetric colliding pairs with
symmetry parameter $A_{s}$=$\left(\frac{N-Z}{A}\right)$ = $0.0$
only. Here, N and Z belong to the combined neutron and proton
content of the reaction. This comparison, which covers a wide
spectrum, will give us a unique possibility to compare various
modeled proximity potentials. We also plan to modify the proximity
potential and the potential by Denisov and we will show that these
new versions improve the agreement with the experimental data.
Section~\ref{model} describes the Formalism in brief,
Section~\ref{result} depicts the results and Summary is presented
in Section~\ref{summary}.
\section{\label{model} Formalism:}
  According to the original version of the proximity potential 1977~\cite{blocki77}, the interaction potential $ V_{N}(r)$ between two surfaces
 can be written as
 \begin{equation}
V_{N}^{Prox~77}(r)= 4\pi \gamma b \overline{R} \Phi \left( r-C_{1}
-C_{2}/b\right) {~\rm MeV},\label{eq:1}
\end{equation}
 where $\gamma$ is the surface energy coefficient. The mean curvature
radius, $ \overline{R}$ in Eq.~(\ref{eq:1}) is written as:
\begin{equation}
\overline{R} = C_{1}C_{2}/(C_{1}+ C_{2}); ~C_{i}=
R_{i}\left[1-\left(b/R_{i} \right)^{2}+\cdots  \right],
\label{eq:3}
\end{equation}
and ${\rm R_{i}}$, the effective sharp radius, reads as;
\begin{equation}
R_{i}= 1.28A^{1/3}_{i}- 0.76+0.8A^{-1/3}_{i} {~\rm
fm}~~~~~~~~~~~(i=1,2). \label{eq:5}
\end{equation}
 This model is referred to as Prox 77, and the corresponding
potential is referred to as $ V_{N}^{Prox~77}(r)$.

Later, Reisdorf \cite{wr94} modified the preceding potential with
a different $\gamma$ value. This is labeled as Prox 88.

The latest version of the proximity potential by Myers and
\'Swi\c{a}tecki (labeled here as Prox 00) is given in
Ref.~\cite{ms2000}. This Prox 00 uses an experimental value and/or
an old formula for the radius.
\par
Recently, Royer and Rousseau~\cite{gr09} gave a more precise
radius formula
\begin{equation}
R_{i}= \alpha A_{i}^{1/3} \left\{1+\beta/A_{i}-\delta
A_{si}\right\} {~\rm fm}~~(i=1,2), \label{eq:8}
\end{equation}
where $\alpha$, $\beta$, and $\delta$  are constants that have
values of 1.2332, 2.348443 and 0.151541, respectively. This
formula is obtained by analyzing as many as 2027 masses with N, Z
$\geq$ 8 and a mass uncertainty $\leq$ 150 keV.  We implement this
radius formula in Prox 00 and label it as Prox 00DP.

Based on the proximity concept, many other potentials have also
been shown in the literature. We will use the potentials by Bass
1973~(labeled here as Bass 73)~\cite{bass73}, similarly, Bass
1977~(Bass 77)~\cite{bass77}, Bass 1980~(Bass 80)~\cite{wr94},
Christensen and Winther 1976~(CW 76)~\cite{cw76}, Broglia and
Winther 1991~(BW~91)~\cite{wr94}, Aage~
Winther~(AW~95)~\cite{aw95}, Ng\^o~1975~ (Ng\^o 75)~\cite{ngo75},
Ng\^o~1980~ (Ng\^o 80)~\cite{ngo80}, and Denisov~\cite{deni02}.

The potential by Denisov~\cite{deni02} is also modified here to
include the previous more precise radius formula [Eq.
(\ref{eq:8})] in its parametrization. This is labeled as Denisov
DP.

The exact potential based on the Skyrme energy density
formalism~\cite{rkp1} (labeled as EDF Exact) along with its
parametrized form (labeled as EDF Par)~\cite{rkp92} will also be
used for comparison. Note that Skyrme forces are also widely used
in intermediate energies~\cite{sk}.

%%%%%%%%%%%%%%%%%%%%%%%%%%%%%%%%%%%%%%%%%%%%%%%%%%%%%%%%%%%%%%%
\section{\label{result}Results and Discussions}
 In Fig. 1, we display the nuclear part of the interaction
 potential $V_{N}(\rm MeV)$ as a function of internuclear distance r~(in femtometers)
using some of the previously listed versions of the proximity
potentials for the reactions of $^{16}$O+$^{16}$O and
$^{40}$Ca+$^{40}$Ca. In Fig. 1(a), we display three versions of
the potentials by Bass, whereas in Fig. 1(b) we deals with Ng\^o
parametrizations. The potentials of Winther and
 collaborators are displayed in Fig. 1(c),  followed by four
 versions of a proximity potential displayed in Fig. 1(d).
\par
 From the figure, we see that different versions of the Bass potentials as well
 as the CW~76 do not have a repulsive core at shorter distance. On the other hand, the BW~91 and AW~95 potentials, follow the Woods-Saxon-type
 distributions. All other potentials have acceptable
 shape: that is, attractive at long distances followed by repulsive at
 shorter distances. Interestingly, the deepest
 potential caused by the proximity potential Prox 88 (=72 MeV).
In other words, one see a huge  difference in
 the potentials obtained from different versions of proximity-based
 formalisms even within the same model. For instance, different versions
 of the proximity potential differ by as much as 17 MeV for the reaction of  $^{40}$Ca+$^{40}$Ca.
 Since the fusion process is a low-density phenomenon happening at the outer surface, the difference in the inner part of the potential may not be so important.
 By adding the Coulomb potential to the nuclear part, one can compute the total
 potential $V_{T}(r)$ as
 \begin{eqnarray}
 V_{T}(r)= V_{N}(r) + V_{C}(r),\\
         =V_{N}(r) + \frac{Z_{1}Z_{2}e^{2}}{r}.
\label{eq:30}
\end{eqnarray}
 Since the fusion
happens at a distance greater than the touching configuration of
the colliding pair, the previous form of the Coulomb potential is
justified. One can extract the barrier height $V^{theor}_{B}$ and
the barrier position $R^{theor}_{B}$ using the following
conditions
\begin{equation}
\frac{dV_T(r)}{dr}|_{r = R^{theor}_{B}} = 0;~~ {\rm{and}} ~~
\frac{d^{2}V_T(r)}{dr^{2}}|_{r = R^{theor}_{B}} \leq 0.
\label{eq:31}
\end{equation}
 \par
 In Fig. 2, we display the total interaction (nuclear +
 Coulomb) potential  $V_{T}(\rm MeV)$  as a function of  internuclear distance
$``r''$~(in femtometers).  We display the results using different
versions of the proximity potentials (Prox 77, Prox 88, Prox 00,
and Prox 00DP) and potentials of Winther and collaborators i.e.,
(CW 76, BW 91, and AW 95). From Fig. 2,  we note
 that, although various potentials differ  significantly in the interior
 part, very little dependence is visible on the surface part and also on the barrier region.
 In terms of fusion barrier heights and positions, all
 versions yield nearly the same  barrier heights and positions.
 Of course, the shape and the curvature of the potential  differs, indicating a different picture  for subbarrier
 fusion cross sections that depends  sensitively on the  shape of
 the  potential one is using. For the reaction of $^{40}Ca+^{40}Ca$, no difference is seen between Prox 00 and Prox 00DP.
This happens because the experimental value of the charge
distribution is available for the $^{40}Ca$ nucleus, and both
models use the same experimental values. The results may differ
for those nuclei in which experimental values are not available.
\par
Knowledge of the shape of the potential, as well as the barrier
position and the height, allows one to calculate the fusion cross
section at a microscopic level. To study the fusion cross
sections, we will use the model given by Wong ~\cite{wg72}. In
this formalism, the cross section for complete fusion is given by
\begin{equation}
\sigma _{fus}= \frac{\pi}{k^{2}}\sum _{l=0}^{ l_{max}} \left(2l+1
\right)T_{l}\left(E_{cm} \right), \label{eq:32}
\end{equation}
where $k= \sqrt{2 \mu E/\hbar^{2}}$ and here $\mu$ is the reduced
mass. The center-of-mass energy is denoted by $E_{cm}$. In this
formula, $l_{max}$ corresponds to the largest partial wave for
which a pocket still exists in the interaction potential, and
T$_{\l}\left(E_{cm} \right)$ is the energy-dependent barrier
penetration factor and is given by,
\begin{equation}
T_{\l}\left(E_{cm} \right)= \left\{1+ \exp \left[ \frac{2
\pi}{\hbar \omega_{\l}} \left(V^{theor}_{B_{\l}} - E_{cm}
\right)\right] \right\}^{-1},
 \label{eq:33}
\end{equation}
where $\hbar\omega_{l}$ is the curvature of the inverted parabola.
If we assume that the barrier position and the width are
independent of $\l$, the fusion cross section reduces to
\begin{eqnarray}
\sigma _{fus}(mb)= \frac{10 R^{theor^{2}}_{B}\hbar \omega_{0}
}{2E_{cm}}\times\nonumber~~~~~~~~~~~~~~~~~~~~~~~~~~~~~~~~~ \\
 \ln \left\{1+ \exp\left[\frac{2\pi}{\hbar \omega _{0}}
\left(E_{cm}-V^{theor}_{B} \right)\right] \right\}. \label{eq:34}
\end{eqnarray}
For E$_{cm}$$>>$V$^{theor}_{B}$, the preceding formula reduces to
a well-known sharp cutoff formula
\begin{equation}
\sigma _{fus}(mb)= 10 \pi R^{theor^{2}}_{B} \left(1 -
\frac{V^{theor}_{B}}{E_{cm}} \right), \label{eq:35}
\end{equation}
whereas for E$_{cm}$$<<$V$^{theor}_{B}$, the foregoing formula
reduces to
\begin{equation}
\sigma _{fus}(mb)= \frac{10 R^{theor^{2}}_{B}\hbar \omega_{0}
}{2E_{cm}}\exp\left\{\frac{2\pi}{\hbar \omega _{0}}
\left(E_{cm}-V^{theor}_{B} \right)\right\}. \label{eq:36}
\end{equation}
We used Eq.~(\ref{eq:34}) to calculate the fusion cross sections.
\par
In Fig. 3, we display the theoretical barrier heights $
V_{B}^{theor}(\rm MeV)$ verses the experimental barrier heights
$V_{B}^{expt}(\rm MeV)$ using all 16 different potentials. The
experimental barrier heights $V_{B}^{expt}$ are taken directly
from the literature
directly~\cite{Vaz81,Aljuwair84,barreto83,Gary82,Morsad90,tomasi82,Trotta01}.
The limited numbers of reactions in certain cases are caused by
the restrictions posed on different
potentials~\cite{rkp1,deni02,ngo75,ngo80,rkp92}.  The lines are
the fits  over the points. These fitted equations distinguish tell
the deviation from the experimental data. Very
 interestingly, we see that, all models
 can reproduce the experimental barrier heights for symmetric colliding nuclei within
$\pm$8\% on average. We also notice that, on average, barriers
formed using
 $\rm EDF ~Exact$, $\rm EDF ~Par$, Bass 77,  Bass 80, Denisov DP,
 and the different versions by
Winther and collaborators are close to the experimental data. On
the other hand, barriers formed within Bass 73, Ng\^{o} 80 and
Prox 77 potentials deviate by $\pm$ 7\% from the experimental
values. The revised versions by Bass improve the barrier heights
drastically. Now Bass 77 and Bass 80 reproduce the
 experimental data within $ 1.5\%$.  A newer version of the Bass potential (Bass
80) shows slight improvement over Bass 77. Strangely, the Ng\^o 80
version deviates more than $7\%$, whereas its first version, was
able to reproduce the barrier heights within $ 2\%$.
 For Ng\^{o} 75, only two systems fall within
 its parametrization limits.

\par
 The four versions of the proximity
potentials yield an interesting comparison. As pointed out by
various authors~\cite{ms2000}, Prox 77 deviates from the
experimental fusion barrier heights by 4\%~(here it is 6.73\%)
whereas, the new versions of the proximity results are lowered to
0.01\%~(here, it is 5.34\%). Note that in Ref. \cite{ms2000}, very
old data for the fusion barriers were used. In this paper, we have
used the latest data. It is worth mentioning that a slight
variation in the radius formula (Prox 00 and Prox 00DP), can
improve the comparison by nearly 1\%. Very interestingly, a change
in the value of the surface energy coefficient $\gamma$ (Prox 77
and Prox 88) improves the agreement drastically~\cite{id}.
Further, the advantage of a new proximity potential (Prox 00) can
clearly be obtained by just using $\gamma$ in Prox 77 as suggested
by Reisdorf~\cite{wr94}. We do not see a direct advantage of the
original Denisov form of the potential over the Skyrme energy
density model by Puri and Gupta~\cite{rkp92} in which a perfect
comparison with the experimental data is clearly visible. Its new
form, Denisov DP, however, shows perfect agreement with the
experimental data. From these figures and this analysis, it is
very clear that different models do not yield very different
results. Instead, technical parameters such as the surface energy
coefficient $\gamma$ and the radius, can have a significant impact
on the outcome. The implementation of the latest radius formula
clearly yields better agreement.

In Fig. 4, we display the fusion barrier positions as a function
of experimentally extracted values. We see that no trend emerges
in this case. This happens because of a great amount of
uncertainty in the measurement of the fusion barrier positions
reported by various authors in various
experiments~\cite{Aljuwair84,Trotta01}.

In Fig. 5, we display the percentage difference of the fusion
barrier heights over its experimental values defined as;
\begin{equation}
\Delta V_{B}~(\%)= \frac{V_{B}^{theor} -
V_{B}^{expt}}{V_{B}^{expt}}\times 100. \label{eq:49}
\end{equation}
 We see that on an individual bases, all proximity potentials can
reproduce the data within $\pm10$\%. The least amount of deviation
is attained from Bass 77, Bass 80, different Winther potentials
(CW 76, BW 91, and AW 95), EDF Exact, EDF Par and Denisov DP~( all
within $\pm5\%$). However, the proximity potential (Prox 77)
deviates much more compared to its other versions.
\par
In Fig. 6, we display the fusion cross section $\sigma_{fus}$ (in
millibarns) as a function of center-of-mass energy $E_{cm}$ for
the reactions of $^{24}$Mg + $^{28}$Si [Fig. 6((a)] and $^{40}$Ca
+ $^{40}$Ca~[Fig. 6(b)]. Here, the latest versions of the
proximity parametrizations along with the original proximity
potential and its modifications are shown for clarity. The
experimental data are taken from the
Refs.~\cite{Gary82,Morsad90,Aljuwair84,tomasi82,barreto83}. As we
see, Bass 80 Denisov DP, and AW 95 do a better job for the
reaction of $^{24}$Mg + $^{28}$Si whereas Prox 77 and Ng\^o 80
fail to come close to the experimental data. For the reaction of
$^{40}$Ca + $^{40}$Ca, no clear picture emerges. In both cases,
the potential of Denisov DP and AW 95 are able to reproduce the
cross section.

%%%%%%%%%%%%%%%%%%%%%%%%%%%%%%%%%%%%%%%%%%%%%%%%%%%%%%%%%%%%%%%%
\section{\label{summary}Summary}
By using as many as 16 versions of the proximity potential derived
either from the proximity potential or from the parametrized
versions in terms of the proximity concept, we carried out a
comparative study of fusion barriers for symmetric colliding
nuclei. For the present study, four versions of the proximity
potential, three versions of the proximity potential by Bass,
three versions of the proximity potentials by Winther and
collaborators, two versions of the proximity potentials by Ng\^{o}
and two versions of the proximity potentials by Denisov and EDF
each were taken. We also proposed new versions of the proximity
potential and a proximity potential by Denisov.
 A detailed study reveals that all
potentials can reproduce experimental data, on average, within
$\pm8$\%. However, the comparison of fusion cross sections reveals
that the Bass 80 Denisov DP, and AW 95 potentials reproduce data
better than the other potentials.

%%%%%%%%%%%%%%%%%%%%%%%%%%%%%%%%%%%%%%%%%%%%%%%%%%%%%%%%%%%%%%%%
 This work was supported by a research grant from the Department of
Atomic Energy, Government of India.

%%%%%%%%%%%%%%%%%%%%%%%%%%%%%%%%%%%%%%%%%%%%%%%%%%%%%

%%%%%%%%%%%%%%%%%%%%%%%%%%%%%%%%%%%%%%%%%%%%%%%%%%%%%%%%%%%%%%

\newpage

\begin{figure}
\centering
\includegraphics* [scale=0.4]{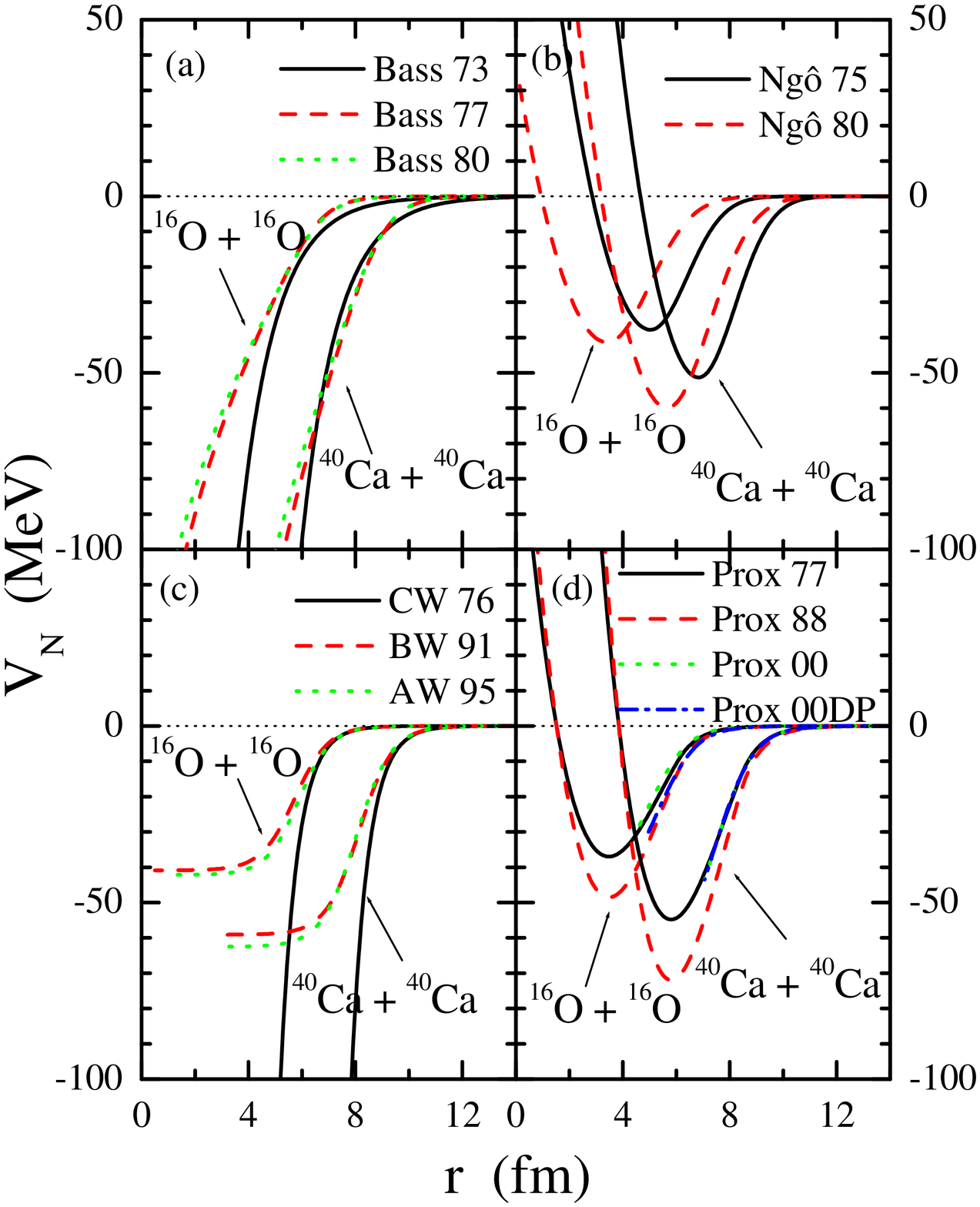}% Here is how to import EPS art
\vskip -0.8 cm \caption {(Color online) The nuclear part of the
interaction potential, $V_{N}(\rm MeV)$ as a function of
internuclear distance $r$~(in femtometers) for the reactions of
$^{16}$O + $^{16}$O and $^{40}$Ca + $^{40}$Ca using different
proximity potentials.} \vskip -0.4 cm
\end{figure}
%%%%%%%%%%%%%%%%%%%%%%%%%%%%%%%%%%%%%%%%%%%%%%%%%%%%%%%%%%%%%%%%
%%%%%%%%%%%%%%%%%%%%%%%%%%%%%%%%%%%%%%%%%%%%%%%%%%%%%%%%%%%%%%%%%%%%
\begin{figure}
\centering
\includegraphics* [scale=0.4]{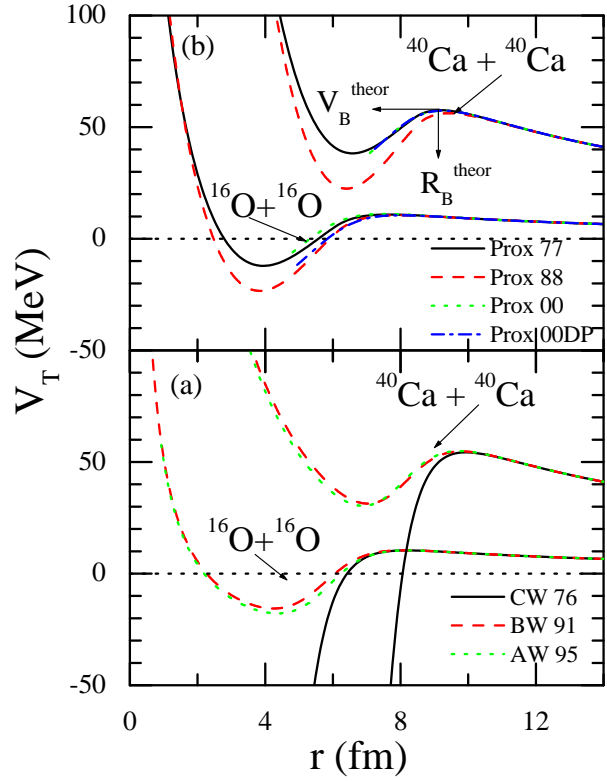}% Here is how to import EPS art
\vskip -0.8 cm \caption {(Color online) The total interaction
potential~$V_{T}(\rm MeV)$ as a function of internuclear distance
$r$~(in femtometers). Here, we display the results obtained with
different versions of proximity and Winther potentials only.}
 \vskip -0.4 cm
\end{figure}
%%%%%%%%%%%%%%%%%%%%%%%%%%%%%%%%%%%%%%%%%%%%%%%%%%%%%%%%%%%%%%%%
%%%%%%%%%%%%%%%%%%%%%%%%%%%%%%%%%%%%%%%%%%%%%%%%%%%%%%%%%%%%%%%%%%%%
\begin{figure}
\centering
\includegraphics* [scale=0.4]{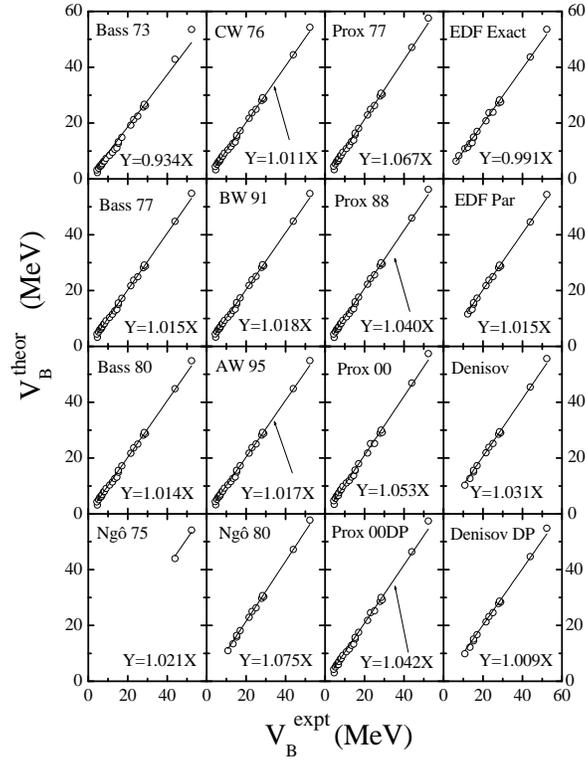}% Here is how to import EPS art
\vskip -1.1 cm \caption {Comparison of theoretical fusion barrier
heights $V_B^{theor}(\rm MeV)$ using different proximity
potentials with experimental values $V_B^{expt}(\rm
MeV)$~\cite{Vaz81,Aljuwair84,barreto83,Gary82,Morsad90,tomasi82,Trotta01}.
The solid lines represent the straight line least squares fit
created over different points.} \vskip -0.4 cm
\end{figure}
%%%%%%%%%%%%%%%%%%%%%%%%%%%%%%%%%%%%%%%%%%%%%%%%%%%%%%%%%%%%%%%%
\begin{figure}
\centering
\includegraphics* [scale=0.4]{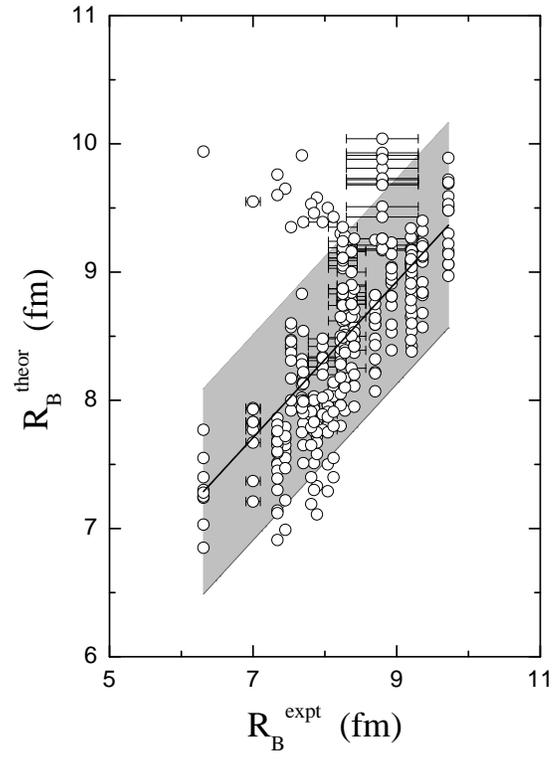}% Here is how to import EPS art
\vskip -0.8 cm \caption {A comparison of theoretical
$R_B^{theor}(\rm fm)$ and experimental fusion barrier positions
$R_B^{expt}(\rm
fm)$~\cite{Aljuwair84,barreto83,Gary82,Morsad90,tomasi82,Trotta01,Vaz81}
using various versions of the proximity potential. Solid line
represent the straight line least square fit.} \vskip -0.4 cm
\end{figure}
%%%%%%%%%%%%%%%%%%%%%%%%%%%%%%%%%%%%%%%%%%%%%%%%%%%%%%%%%%%%%%%%%%%%%%%%%%%%%%%%
\begin{figure}
\centering
\includegraphics* [scale=0.4]{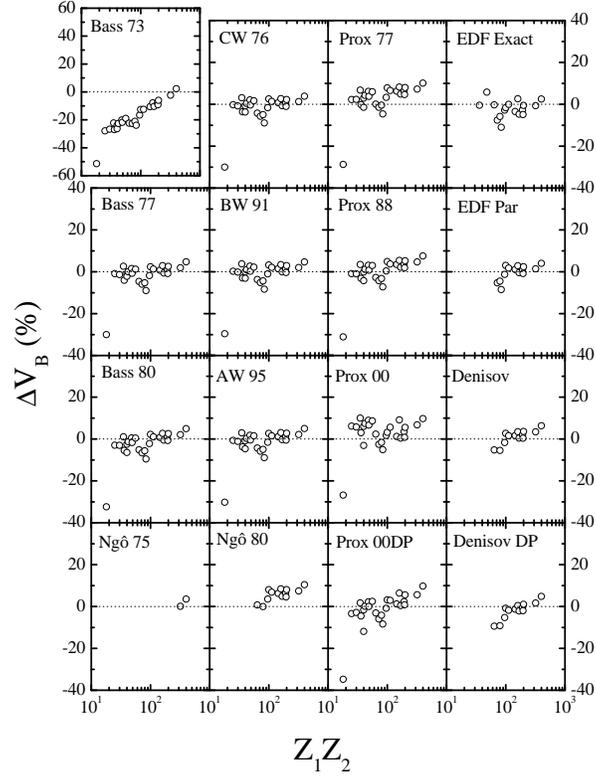}% Here is how to import EPS art
\vskip -0.6 cm \caption {The percentage deviation $\Delta
V_{B}~(\%)$ as a function of the product of charges $Z_{1}Z_{2}$
using different versions of the proximity potential.} \vskip -0.4
cm
\end{figure}
%%%%%%%%%%%%%%%%%%%%%%%%%%%%%%%%%%%%%%%%%%%%%%%%%%%%%%%%%%%%%%%%%%%%%%%%%%%%
\begin{figure}
\centering
\includegraphics* [scale=0.4]{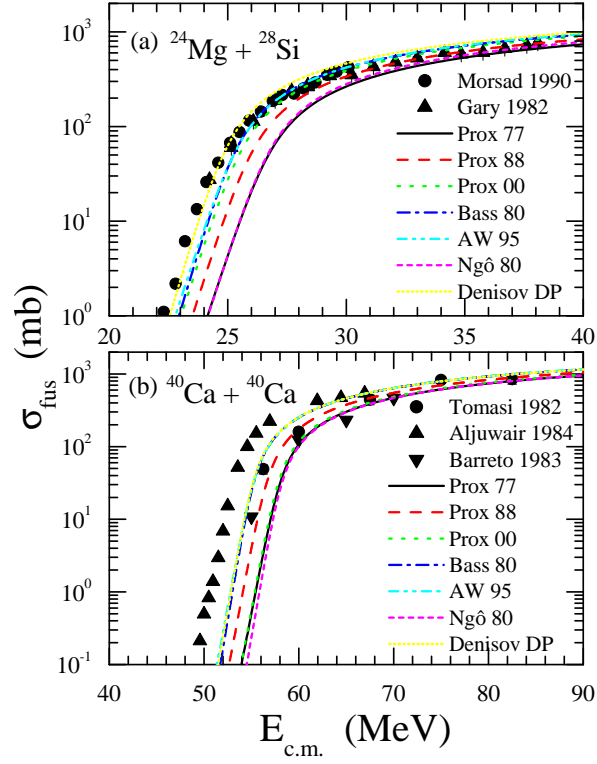}% Here is how to import EPS art
\vskip -0.8 cm \caption {(Color online)The fusion cross-sections
for the reactions of $^{24}Mg+^{28}Si$ [Fig. 6(a)] and
$^{40}Ca+^{40}Ca$ [Fig. 6(b)] as a function of center-of-mass
energy $E_{cm}$. The experimental data are taken from Morsad
1990~\cite{Morsad90}, Gary 1982~\cite{Gary82}, Tomasi
1982~\cite{tomasi82}, Aljuwair 1984~\cite{Aljuwair84}, and Barreto
1983~\cite{barreto83}. For the clarity, only the latest versions
of the different proximity potentials are shown.} \vskip -0.4 cm
\end{figure}
%%%%%%%%%%%%%%%%%%%%%%%%%%%%%%%%%%%%%%%%%%%%%%%%%%%%%%%%%%%%%%%%%%%%%%

\end{document}